\documentclass{raa}  

\usepackage{graphicx,times}
\usepackage{natbib}
\usepackage{amssymb,amsmath}
\bibpunct{(}{)}{;}{a}{}{,}

\usepackage[pagebackref=true]{hyperref}

\begin{document}

   \title{The First Scientific Flight and Observations of the 50-mm Balloon-Borne White-Light Coronagraph}

 \volnopage{ {\bf 20XX} Vol.\ {\bf X} No. {\bf XX}, 000--000}
   \setcounter{page}{1}

   \author{Kaifeng Kang\inst{1,2}, Min Huang\inst{3}, Yang Liu\inst{3}, Jun Lin\inst{1,2,4,5}, Tengfei Song\inst{1,2}, Xuefei Zhang\inst{1,2}, Dayang Liu\inst{6}, Tao Zhang\inst{7}, Yan Li\inst{1,2}, Jingxing Wang\inst{1,2}, Mingzhe Sun\inst{6}, Mingyu Zhao\inst{1,2}, Guangqian Liu\inst{1,2},  Xianyong Bai\inst{8}, Lidong Xia\inst{6}, Yu Liu\inst{9}
   }
   
   \institute{ Yunnan Observatories, Chinese Academy of Sciences,  Kunming, Yunnan 650216, China; {\it jlin@ynao.ac.cn}\\
        \and
             Yunnan Key Laboratory of the Solar physics and Space Science, Kunming, Yunnan 650216, China\\
	\and 
Aerospace Information Research Institute, Chinese Academy of Sciences, Beijing 100094, China\\
\and 
University of Chinese Academy of Sciences, Beijing 100049, China\\
\and Center for Astronomical Mega-Science, Chinese Academy of Sciences, Beijing 100101, China\\
\and
Institute of Space Sciences, Shandong University, Weihai,  Shandong 264209, China\\
\and Kunming University of Science and Technology, Kunming, Yunnan 650500, China\\
\and National Astronomical Observatories, Chinese Academy of Sciences, Beijing 100101, China
\and Southwest Jiaotong University, Chengdu, Sichuan 610031, China\\
\vs \no
   {\small Received 20XX Month Day; accepted 20XX Month Day}
}

\abstract{A 50-mm balloon-borne white-light coronagraph (BBWLC) to observe white-light solar corona over the altitude range from 1.08 $\text{R}_{\odot}$ to 1.50  $\text{R}_{\odot}$ has recently been indigenously developed by Yunnan Observatories in collaboration with Shangdong University (in Weihai) and Changchun Institute of Optics, Fine Mechanics and Physics, which will significantly improve the ability of China to detect and measure inner corona. On 2022 October 4, its first scientific flight took place at the Dachaidan area in Qinghai province of China. We describe briefly the BBWLC mission including its optical design, mechanical structure, pointing system, the first flight and results associated with the data processing approach. Preliminary analysis of the data shows that BBWLC imaged the K-corona with three streamer structures on the west limb of the Sun. To further confirm the coronal signals obtained by BBWLC, comparisons were made with observations of the K-coronagraph of the High Altitude Observatory and the Atmospheric Imaging Assembly on board the Solar Dynamics Observatory. We conclude that BBWLC eventually observed the white-light corona in its first scientific flight.
}

   \authorrunning{K. Kang et al. }  
   \titlerunning{The first flight of balloon-borne coronagraph} 
   \maketitle

%
\section{Introduction}  
\label{sect:intro}

The solar corona is the tenuous outermost atmosphere of the Sun that consists of magnetized plasma of millions of degrees in temperature, extending outward to several solar radii and beyond. On the one hand, much of the coronal plasma is confined by the magnetic field in the form of twisted arcade-like structures and closed loops \citep{2019ARA&A..57..157C}. On the other hand, some coronal plasma expands to fill interplanetary space as a continuous supersonic outflow known as the solar wind \citep{1958ApJ...128..664P}. It can be seen that the corona is closely related to the two most crucial problems in solar and space plasma physics, which  are the corona heating and the solar wind acceleration, respectively \citep{2006LRSP....3....1M, 2025A&A...695A.192C}.

As the region where the above two processes take place, the corona includes key information of the physics of these processes, and the observation of the corona is crucial for understanding the physical processes involved. The white-light observation plays an important role in revealing changes of the plasma density of the corona \citep{2008ApJ...683.1168K, 2015A&A...583A.101M}. Thus, white light images of the corona can be used to constrain the distribution of electron density and morphology of the corona over time \citep{2016JGRA..121.7470T}. In addition, observations of the density structure in the low corona are a crucial constraint on determining the location where the solar wind transitions from sub-Alfv\'{e}nic to super-Alfv\'{e}nic supersonic and on theories of the corona heating \citep{2016JGRA..121.7470T}. In other words, theoretical models of the corona heating and the solar wind acceleration can be constrained by white-light observations of the corona.

In addition, the highly ionized and conducting corona is also closely related to one of the most energetic eruptive phenomena in the solar system, namely, coronal mass ejections (CMEs), which are the primary driver of the disastrous space weather \citep{2002ApJ...576..485L, 2003NewAR..47...53L, 2006SSRv..123..251F, 2011LRSP....8....1C}. CMEs are usually identified as large-scale disturbances in the coronal intensity-image sequences and thus can be unambiguously detected and tracked in white light \citep{2013SoPh..288..637T}. Therefore, many well-known properties of CMEs come from white light observations \citep{1976SoPh...48..389G, 1993JGR....9813177H, 1999JGR...10412493S, 2012SPIE.8444E..3ND}. For example, white light observations provide  data on the basic properties of CMEs such as size, speed, acceleration and mass. In turn, these discoveries further strengthened the need for white light observations, particularly in the region of the inner corona. As a consequence, observing the corona in white light will not only advance our understanding of the CMEs initiation, acceleration and propagation, but also improve our capability of forecasting the disastrous space weather.

However, before the French astronomer Bernard Lyot invented the coronagraph in the 1930s, the corona could only be observed during a total solar eclipse since the brightness of the corona is around millionth that of the solar disk \citep{1939MNRAS..99..580L}. Even with a coronagraph, it is also difficult to observe the corona in the white-light bandpass (i.e., the white-light corona, WLC) on the ground because the scattering light of the Earth atmosphere could still be much stronger than the signal of WLC. So good observatory site is needed for observing the corona, at which the scattering of the Earth atmosphere is at low level. Usually, this kind of site is located at high mountain where the atmosphere is tenuous and scattering could be weak \citep{1998EP&S...50..493K, 2008JASTP..70..356S}.

According to the source of coronal light, the WLC is divided into three components, namely, the `kontinuierlich' (continuum) corona (K-corona) resulting from the scattering of the white light of the photosphere by free electrons, emission corona (E-corona) being the emission from ions of high charge state in the corona, and Fraunhofer corona (F-corona, also known as the zodiacal light) resulting from scattering of the photospheric light by the dust around the ecliptic plane \citep{1999PASJ...51..269S, 2008JASTP..70..356S, 2019Natur.576..232H}. Because of the very high temperature of the coronal plasma, its radiation is mainly concentrated in the ultraviolet and X-ray bands, thus the E-corona in the white-light bands is extremely weak. Comparing with the K- and F-coronae, the E-corona makes up a very small portion of the WLC brightness and is usually negligible \citep{2008JASTP..70..356S}. For this reason, the WLC is usually called the scattered corona which consists of K- and F-coronae. One of the differences between K-  and F-coronae is that the former is linealy polarized while the latter at heights less than about 4 $\text{R}_{\odot}$ is unpolarized \citep{2001ApJ...548.1081H}. This is particularly important for observations of the K-corona. For instance, the only ground-based white light coronagraph, the K-coronagraph (K-Cor) at the Mauna Loa Solar Observatory (MLSO), observes the K-corona by measuring the linear polarization brightness of the WLC \citep{2012SPIE.8444E..3ND, 2016JGRA..121.7470T}.

In addition to ground-based observations of WLC, the space-borne observation is the most common choice. This is because it  not only avoids the influence of the scattering light from the Earth atmosphere, but also obtains images close to the diffraction limit. This type of observation is typified by the coronagraphs aboard the Orbiting Solar Observatory-7 (OSO-7;  \citealt{1975ApOpt..14..743K}), the Skylab \citep{1974JGR....79.4581G}, the Solar Maximum Mission (SMM; \citealt{1981ApJ...244L.117H}), the Solar and Heliospheric Observatory (SOHO; \citealt{1995SoPh..162..357B, 2004JGRA..109.7105Y}) and the Solar Terrestrial Relations Observatory (STEREO; \citealt{2003SPIE.4853....1T}) since the 1970s, which have provided numerous discoveries associated with  WLC. Moreover, another option for observing the WLC is the balloon-borne observation (hereafter referred to as BBO) that is the stratospheric observation made in the near space by using balloon \citep{2024ChJSS..44.1068L}. The BBO has the following several major advantages.

First of all, comparing with the ground-based observation, the BBO is only slightly affected by scattering light of the Earth atmosphere, which is  favorable for observations of the WLC. Second, it can detect and measure WLC at a small fraction of the cost of its space-borne counterparts \citep{2011SoPh..268....1B}. Meanwhile, the BBO allows for the recovery, modification, upgrading and reuse of the coronagraph and the related equipments, which is also important for the development, validation and improvement of  techniques for detecting WLC. Due to these advantages and the rich potential of the BBO, a number of projects have been devoted to solar observations from balloon-borne platforms at the stratospheric height (e.g., \citealt{1959ApJ...130..345S, 1972SoPh...26..305K, 1978PASJ...30..337H, 1979SoPh...63...35H, 1996SoPh..164..403R, 2004AdSpR..33.1746B} ). However, because of technical difficulties, only one successful coronal observation experiment has been carried out on a floating platform. That is the Balloon-borne Investigation of Temperature and Speed of Electrons (BITES) mission launched from the Fort Sumner, New Mexico of the United States on September 18, 2019 \citep{2021SoPh..296...15G}.

But only filtergrams of the corona were obtained by BITES, and no observation of the white-light corona has ever been performed in the near space until the 50-mm balloon-borne white-light coronagraph (BBWLC) was developed by the Yunnan Observatories in collaboration with Shangdong University (in Weihai) and Changchun Institute of Optics, Fine Mechanics and Physics. Three ground-based experiments of the BBWLC were conducted at the Gaomeigu Station in Lijiang, Yunnan \citep{2018SoPh..293....1Z} and the Wumingshan mountain in Daocheng, Sichuan \citep{2018SPIE10704E..22L, 2020RAA....20...85S, 2021MNRAS.505.3070S, 2021SPIE12070E..0BL} before the balloon launch, respectively (see \citealt{2023SSPMA..53y9611J} for more details). Here, we will describe in detail its first scientific flight, the data collected, and the consequent results. Section \ref{sect:Cor} presents an overview of the BBWLC including its optical design, mechanical structure and stray light detection. Section \ref{sect:BBCF} describes the BBWLC system and its first scientific flight. Section \ref{sect:res} displays data processing and the  results. Finally, discussions and conclusions are given in Section \ref{sect:discussion} , respectively.

\section{Coronagraph}
\label{sect:Cor}


BBWLC was designed as a traditional internally occulted white-light coronagraph that images the K-corona in 5500 \AA \ with a passband of 50 \AA \ and a field of view (FOV) covering the altitude range from 1.08 $\text{R}_{\odot}$ to 1.50  $\text{R}_{\odot}$ of helieocentric distance. An advantage of the internally occulted coronagraph is that it allows the inner edge of FOV of the coronagraph to be as close to the limb of the Sun as possible in order to observe the inner corona. On the other hand, its disadvantage also exists such that the stray light from the direct sunlight is strong. This places higher demands on the stray light suppression design for the BBWLC.

\begin{table*}
\begin{center}
\caption{Parameters of the BBWLC}

\begin{tabular}{c|c}

\hline 
Parameter    &Value  \\

   \hline
Optical resolution  &$2.8^{\prime\prime}$   \\

Objective lens aperture    &50 mm   \\
 
Inner FOV cutoff  &1.08 $\text{R}_{\odot}$  \\

Outer FOV cutoff    &1.50 $\text{R}_{\odot}$  \\

Wavelength   &5500 \AA   \\

Passband   &50 \AA   \\

Effective focal length   &1587 mm   \\

Scattering light level    &2$\times$10$^{-6}$ $\text{B}_{\odot}$ \\

Weight   &$\sim$25 kg   \\

Size   & 1900 mm$\times$210 mm$\times$180 mm   \\

Detector  &sCMOS, 2048$\times$2048 pixels  \\
\hline
\end{tabular}

\label{tab1}
\end{center}
\end{table*}

\begin{figure}[!htbp]
\centering
	\includegraphics[width=1.0\textwidth]{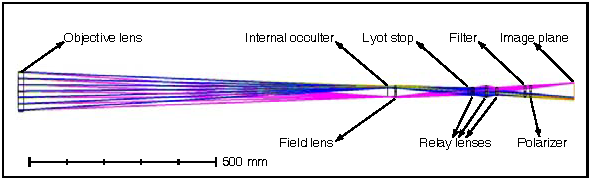}
    \caption{Light path of BBWLC. Sunlight enters the coronagraph from the left.} 
    \label{figure1}  
\end{figure}

\subsection{Optical System Design of BBWLC}
\label{sect:Optic}

Table \ref{tab1} lists the main parameters of the BBWLC and the corresponding optical layout with the sunlight entering the system from the left being shown in the Figure \ref{figure1}. In order to effectively improve image quality, reduce the number of lenses and decrease stray light level (SLL), the objective lens with aperture of 50 mm is designed to be aspherical. It  firstly focuses the direct light from the Sun on the internal occulter where the light from the disk is blocked. In the mean time, the field lens and the internal occulter determine the FOV of the BBWLC. The inner and outer FOV cutoffs from the center of the Sun are 1.08 $\text{R}_{\odot}$ and 1.50 $\text{R}_{\odot}$, respectively. In addition, the diffracted light from the objective lens is imaged onto the Lyot stop, and is thus blocked. Then, the output beam from the Lyot stop is refocused by the relay lenses on the image plane where a sCMOS of 2048$\times$2048 pixels is located. Meanwhile, a filter with the central wavelength of 5500~\AA \ and a passband of 50 \AA \ is used to perform obervations of the corona in white-light. Moreover, in order to observe the K-corona, a polarizer is used for linear polarization observations to filter out non-polarized information of F-corona \citep{2001ApJ...548.1081H}. With such an arrangement, an effective focal length of 1587 mm and an optical resolution of $2.8^{\prime\prime}$ are realised by the optical design.


\subsection{Mechanical Structure Design of BBWLC}
\label{sect:Mech}

\begin{figure}[!htbp]
\centering
	\includegraphics[width=1.0\textwidth]{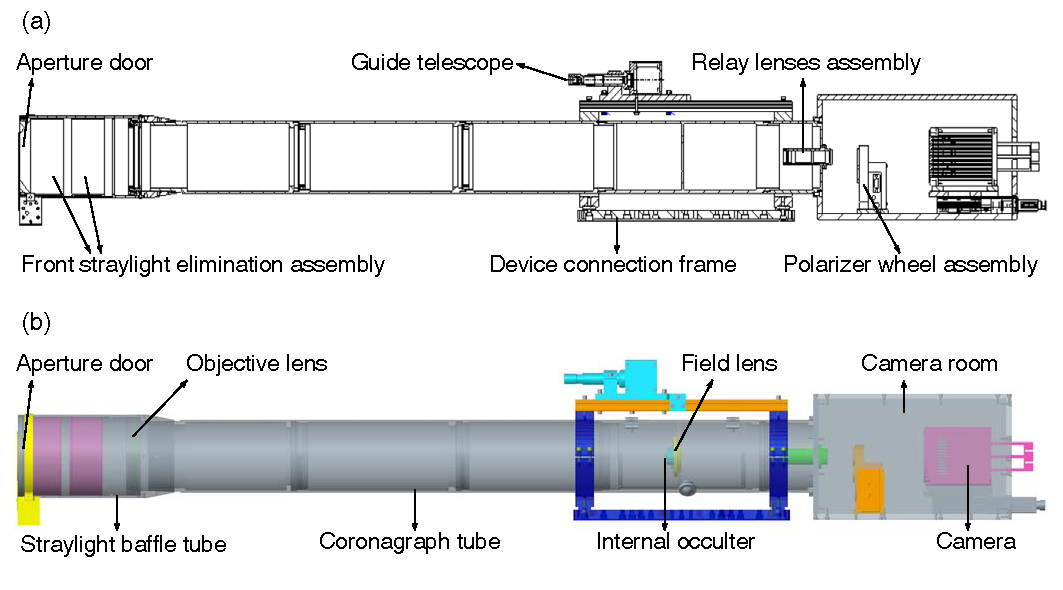}
    \caption{(a) Two-dimensional diagram of the mechanical structure of the BBWLC. (b) The corresponding three-dimensional perspective view of BBWLC.} 
    \label{figure2}  
\end{figure}

The mechanical structure of the BBWLC consists of the following primary components: a coronagraph tube, assembly for the front straylight elimination, aperture door, objective lens (namely primary lens), straylight baffle tube, internal occulter, field lens, relay lens assembly, polarizer wheel assembly, camera room, camera and a device connection frame where a guide telescope is mounted  (see Figures \ref{figure2}(a) and \ref{figure2}(b)). As shown in Figures \ref{figure2}(a) and \ref{figure2}(b), the BBWLC is designed as an inner and outer double-layer cylindrical structure. The coronagraph tube and straylight baffle tube constitute the outer supporting structure that provides a stable and reliable “skeleton” to ensure the rigidity and strength of the whole coronagraph. The objective lens assembly, field lens assembly and relay lens assembly make up the inner structure for mounting each optical component at the correct position as designed. The overall dimensions and the total mass of the mechanical structure are 1900 mm $\times$ 210 mm $\times$ 205 mm and about 25 kg, respectively.

Since dust contamination of the primary lens is the dominant source of the scattering light in an internally-occulted coronagraph \citep{2006Nelson}, the objective lens needs to be cleaned frequently. So, frequent removals and installations  of the primary lens and its cleanliness should be considered when designing the structure of the objective lens assembly. To resolve this problem, a separate mirror base is designed for the objective lens, which can prevent contact with the surface of the primary lens to ensure its cleanliness when removing and installing. Finally, a polarizer wheel assembly is mounted in front of the camera for measuring the polarization of the light. The polarizer wheel has six holes, four of which are for polarized images, one for  unpolarized images, and the sixth for the dark field. Holes 1 through 4 are equipped with four polarizers of 0\textdegree, 90\textdegree, 45\textdegree and 135\textdegree, respectively. 



\begin{figure}[!htbp]
\centering
	\includegraphics[width=0.86\textwidth]{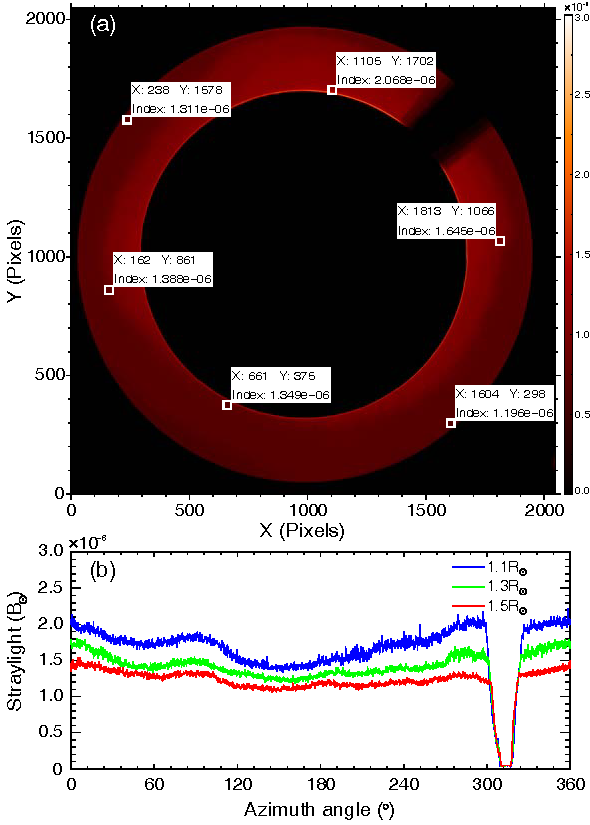}
    \caption{(a) Image for the results of inspecting stray light of the BBWLC before its launch. (b) Variations of the stray light strength in units of 10$^{-6}$ $\text{B}_{\odot}$ versus the azimuth at three locations of 1.1, 1.3 and 1.5 solar radii from the center of the Sun. The sudden drop between azimuths of 300$^{\circ}$ and 320$^{\circ}$ results from the block of the light by the supporter of the occulting disk.} 
    \label{figure3}  
\end{figure}

\subsection{Stray Light Detection}
\label{sect:Str}

On the basis of the above optical system and mechanical structure, the BBWLC achieves a SLL of 2$\times$10$^{-6}$ $\text{B}_{\odot}$ ($\text{B}_{\odot}$ is the mean brightness of the solar disk in the wavelength of 5500~\AA) at 1.10 $\text{R}_{\odot}$. This meets the stray light metrics of the coronagraph required by the project such that the upper limit of the SLL is 10$^{-5}$ $\text{B}_{\odot}$ at 1.10 $\text{R}_{\odot}$.

This is the results from the tests in laboratory, and does not necessarily mean that the level of SSL could keep unchanged afterward. On the other hand, the primary lens is the major source of the scattering light in the coronagraph system, and its basic  property and the working state are essential for the overall quality of the coronal image. Therefore, it is necessary to monitor the status of the primary lens at any time. The scattering sources of the objective lens mainly include dust contamination, imperfect surface polishing,  impurities, and non-uniformity of the internal material of the glass. The scattering light caused by the last three sources is fixed and usually do not change after the lens has been made \citep{1985OptEn..24..380B}. Thus, to monitor the SLL of BBWLC is actually to detect scattering lights caused by dust contaminants on the objective lens.

Because of the disorder and randomness of the dust accumulation on the lens, the scattering light caused by dusts on the objective lens surface is difficult to quantify accurately. This issue had been solved by  \citet{2024RAA....24b5020L} and \citet{Liu2025} via monitoring the scatttering stray light from dusts. The method mainly involves two steps. First, the level of the stray light from the ghost image of the objective lens is simulated and measured. Second, the flux ratio of scattering light and ghost image on the conjugate plane is determined. The flux ratio remains constant on the image plane while it varies with the accumulation of dusts on the lens surface. Consequently, the SLL of dust on the image plane can be quantified by using this ratio combining with the level of ghost image stray light. Using the objective lens with a large number of different surface cleanliness levels, the accuracy of the above method could be verified in the Space Optics Laboratory (SOL) at Shandong University (Weihai, China). The laboratory is a thousand-level ultra-clean darkroom that can realize 20-m optical path without bending, and dedicated to the coronagraph development and test. At the same time, it has favorable conditions for the stray light detection of the coronagraph so that the stray light detection data is reliable.

To ensure that the SLL of the coronagraph meets the observation requirements, the stray light detection needs to be  performed before launching the BBWLC. When monitoring the stray light, two preparations need to be made. First, the objective lens of the BBWLC was thoroughly cleaned to minimize optical contamination from dust on the lens surface. Second, the optical components of the coronagraph, such as field lens, relay lenses, filter and detector, were thoroughly cleaned to ensure that stray light within the optical system was effectively suppressed.

Figure \ref{figure3} (a) displays the image for the results of inspecting stray light while Figure \ref{figure3} (b) shows the variation of the stray light strength in units of 10$^{-6}$ $\text{B}_{\odot}$ versus the azimuth at the locations of 1.1, 1.3 and 1.5 solar radii from the center of the Sun. The sudden drop between azimuths of 300$^{\circ}$ and 320$^{\circ}$ results from the block of the light by the supporter of the occulting disk. Figures \ref{figure3} (a) and \ref{figure3} (b) show that the maximum value of stray light level of the coronagraph is about $2.3 \times$10$^{-6}$ $\text{B}_{\odot}$ in the inner FOV and $1.5 \times$10$^{-6}$ $\text{B}_{\odot}$ in the outer  FOV. Therefore, it can be judged that the overall SLL of the coronagraph is better than $2.3 \times$10$^{-6}$ $\text{B}_{\odot}$, which meets the SLL required by the project.

\begin{figure}[!htbp]
\centering
	\includegraphics[width=0.86\textwidth]{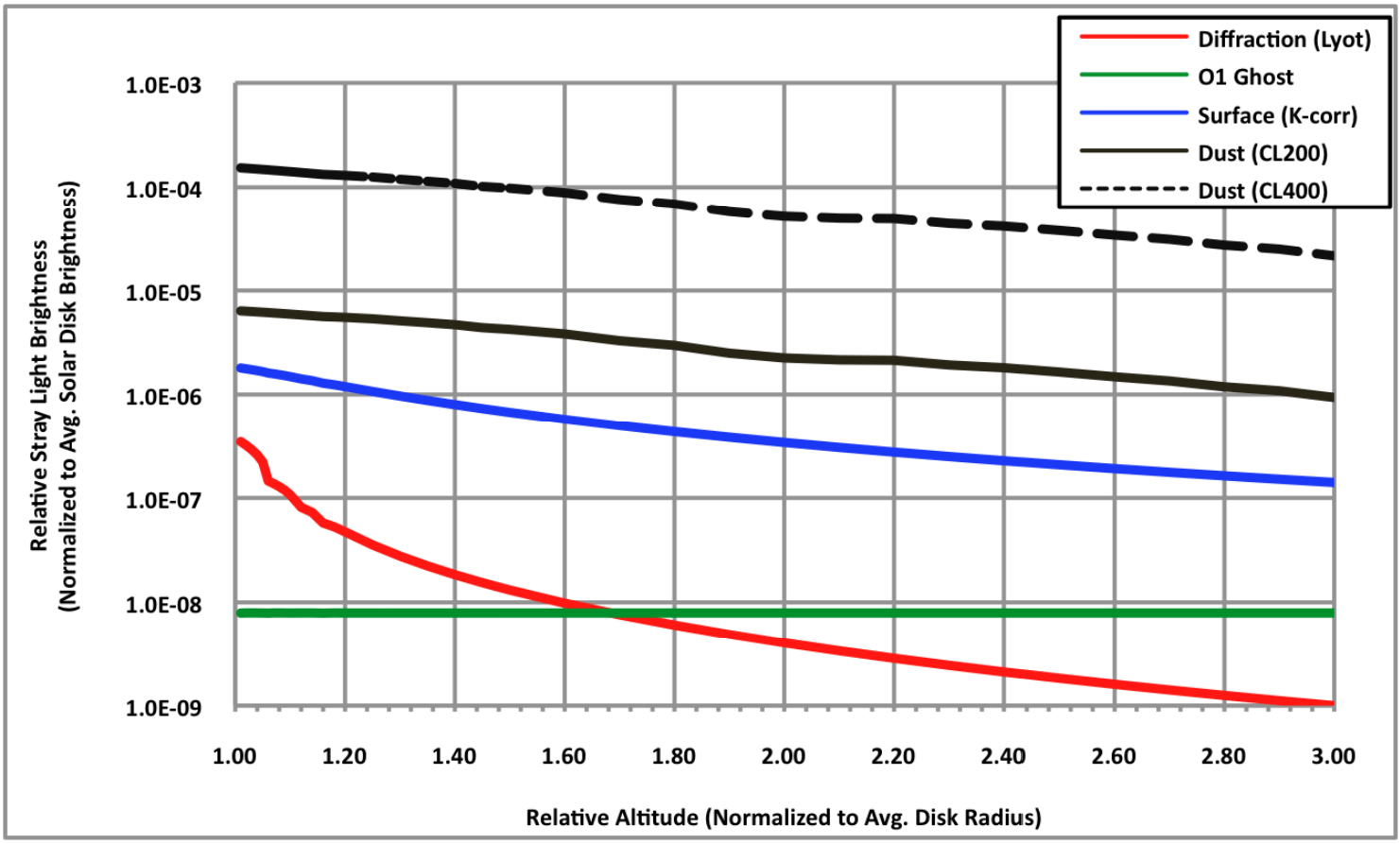}
    \caption{The predicted stray light contributions from diffraction, O1 ghost, surface roughness, and surface contamination. From \citet{2012SPIE.8444E..3ND}.} 
    \label{figureK}  
\end{figure}

For comparison, Figure \ref{figureK} displays a set of simulated results for variations in the brightness of the stray light versus the heliocentric distance in various environments \citep{2012SPIE.8444E..3ND}. The stray light brightness of the K-Cor is also included in the figure. The dashed curve shows an exaggerated case, which might not be realistic. The curve for the K-Cor indicates that levels of the stray light brightness are about $2\times 10^{-6}$~B$_{\odot}$ at 1.1~R$_{\odot}$, and $8\times 10^{-7}$~B$_{\odot}$ at 1.5~R$_{\odot}$, respectively. This indicates that the K-Cor is overall at a good shape.


\section{BBWLC System and Its First Flight}
\label{sect:BBCF}

\subsection{BBWLC System}
\label{sect:BBCS}

\begin{figure}[!htbp]
\centering
	\includegraphics[width=0.7\textwidth]{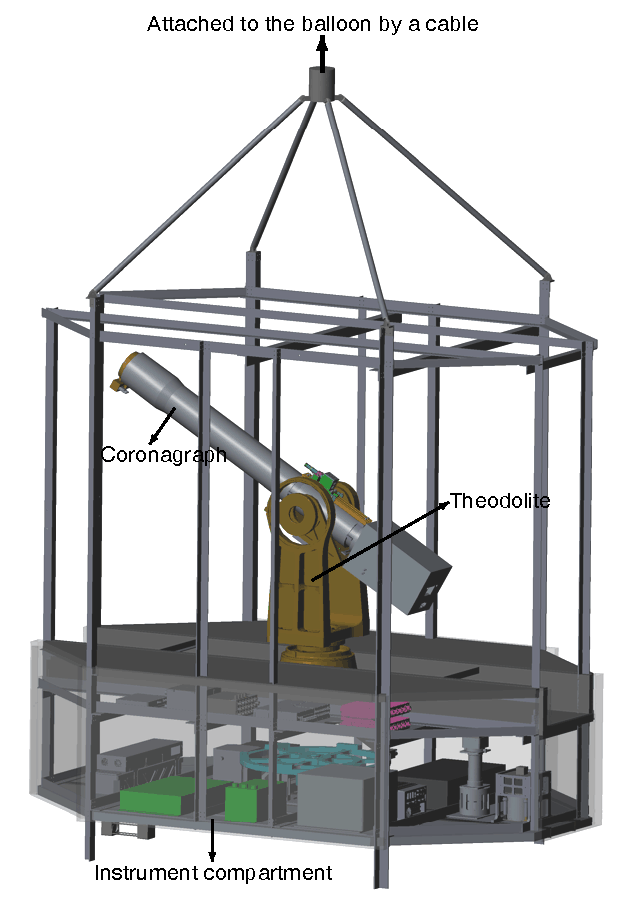}
    \caption{The sketch depicting the basic structure of the BBWLC system that includes a gondola, theodolite and coronagraph.} 
    \label{figure4}  
\end{figure}

As shown in Figure \ref{figure4}, the BBWLC system mainly consists of three components: the coronagraph itself, a theodolite (namely secondary control platform, SCP) and a gondola (namely the first control platform, FCP) that is an aluminium framework, making it relatively light in weight while providing the required stiffness. The coronagraph is mounted on the SCP, which is mounted on the FCP. In order to find the Sun and to point to it, the BBWLC must be free to turn both horizontally and vertically. This is realized by the joint rotation of SCP together with FCP. First of all, FCP finds the Sun and aligns the coronagraph approximately to the Sun within $\pm$1\textdegree \ by means of a solar sensor mounted on the front of the gondola. Then, SCP makes the coronagraph point precisely to the Sun within  $10^{\prime\prime}$ of accuracy using the position information of the Sun center provided by the guide telescope (See figure \ref{figure2}). 


We note here that the proper functioning of the whole BBWLC system, including the rotation of SCP and FCP, depends on electrical power provided by a battery combo, which is composed of a 1.6 kWh battery and a 5.2 kWh battery (see the two green components in Figure \ref{figure4}). The battery combo can provide at least 10 hours of electricity for the BBWLC system and are installed in an instrument compartment at the bottom of the gondola (Figure \ref{figure4}). In addition to the battery combo, the other auxiliaries of the BBWLC system such as communication equipments for communications between the ground and the balloon, flywheel (see the light blue component in Figure \ref{figure4}) that rotates the gondola azimuthly, and industry-level personal computer (IPC; See the pink component in Figure \ref{figure4}) that remotely controls the BBWLC system, are also mounted in the instrument compartment to protect them from accidental damage during flight.

\begin{figure}[!htbp]
\centering
	\includegraphics[width=0.6\textwidth]{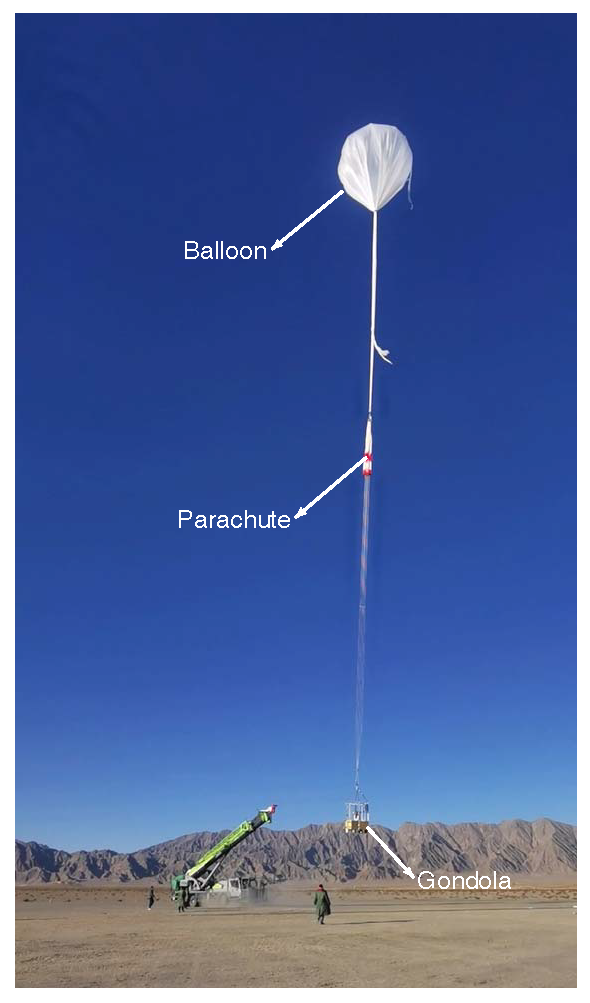}
    \caption{The BBWLC system immediately after launch. The parachute for landing is part of the flight train connecting the gondola payload to the balloon.} 
    \label{figure5}  
\end{figure}

In addition to the guidance, the thermal protection is another important issue that we need to deal with. Since the gondola was open and fully exposed to the ambient air with temperature of $\sim$ $-50$ \textdegree C in the stratosphere during the flight, and the coronagraph tube is made of hard aluminum alloy that is a good thermal conductor, the temperatures inside and outside the BBWLC will soon be about the same. As a result, the performance of the optical components inside the tube of the BBWLC will be seriously affected. In order to ensure the normal operation and image quality of the coronagraph in an environment of  very low temperature condition a thermal protection system is needed to keep temperature of inside the coronagraph tube at $-$5 \textdegree C $\pm$ 3 \textdegree C  \citep{2023SSPMA..53y9611J}. This can be done as follow: several groups of electric heating plates are mounted directly on the outside surface of the coronagraph tube as per the design specifications with electric wires neatly routed. Then, several pieces of the heat-insulating wool are wrapped outside the heating plates. At the third step, the outer side of the heat-insulating wool was wrapped with Aluminum film, which not only prevents inside heat from escaping, but also reflects sunlight to keep the temperature inside as constant as possible.

\subsection{First Flight}
\label{sect:FF}

After each individual component of the BBWLC system was successively tested as separate modules on the ground, the whole system was transported to the launch site at Dachaidan that is located in the northwest of Qinghai Province, the northern edge of the Qaidam Basin. By October 2, 2022, all the apparatus had been assembled and tested. Starting at approximately 02:00 a.m., Beijing time, on October 4, 2022, the final check-out was completed, the BBWLC system was installed on the gondola, and the launching crew went into action. They hooked the gondola onto the crane, and integrated the gondola with the balloon system that includes a balloon, a parachute, and several cables as shown in Figure \ref{figure5}. At 05:00 a.m., all the connections, including cables and electric wires, were completed, and the whole system was waiting for launch.

The helium started to be pumped into the balloon at 07:00 a.m., the launch is ready at 08:30 am. The day is sunny, and suitable for launch. The helium-filled balloon lifted off as the crane released the payload module at 08:52 a.m. eventually. Figure \ref{figure5} shows the gondola, parachute and balloon immediately after the launch. Then, tracking of the balloon began. After ascending for almost 1.5 hours, The BBWLC was successfully sent into the stratosphere at an altitude of 30 km. It drifted westward at a mean speed of 30~km~h$^{-1}$. After a period for testing and debugging the communication equipment, the pointing mechanism (PM) consisting of the FCP and SCP centered the coronagraph on the Sun by remote control. Subsequently, the aperture door of the coronagraph was opened successfully at about 05:00~UT. 

\begin{figure}[!htbp]
\centering
	\includegraphics[width=0.8\textwidth]{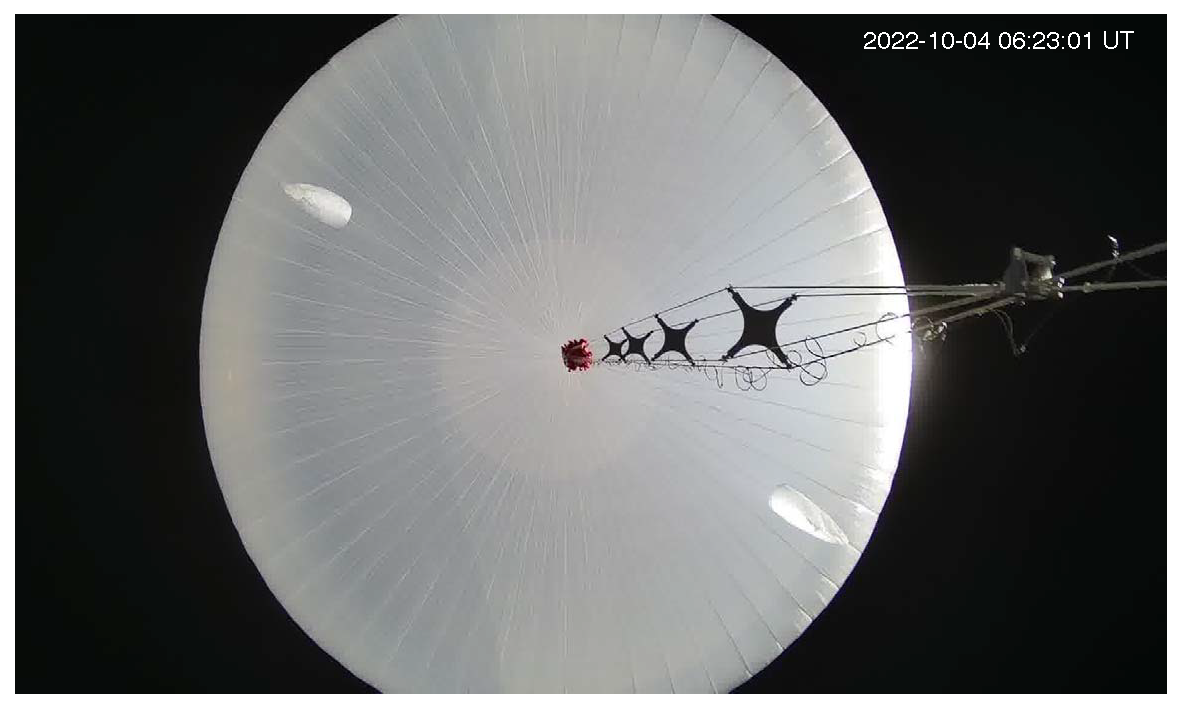}
    \caption{A snapshot of the surveillance video showing that the sky is very clean.} 
    \label{figure9}  
\end{figure}

At 05:02 UT, the camera was turned on after the focus of the telescope had been adjusted first. The observations started and all the data were recorded in the flash storage on board. Figure \ref{figure9} shows a snapshot of the surveillance video during the flight. We can see that the sky at 30~km above the sea level is very clean. During the observation, all onboard electronics, such as the battery pack, the IPC and their associated software, operated normally during most of the mission with the exception of polarizer wheel. Due to a mechanical failure of the polarizer wheel, only the images of polarization at angles of 0\textdegree, 45\textdegree and 90\textdegree \ were obtained, no image of polarization at 135\textdegree. Toward the end of the observation, the flat-field and dark-field of the instrument were obtained at 06:48:51 UT. The observations were stopped at 06:53:37 UT. Flight termination command was issued at 07:00 UT on 4 October 2022, followed by cut-away of the balloon. Then, the gondola descended with the parachute and reached the ground relatively softly. The gondola could serve to protect the payloads, to some extent,  when it hits the ground in addition to serving as a mounting. After retrieved, the BBWLC and the auxiliary equipments were found to suffer from light damage. The payloads were recovered and all the scientifc data stored in the onboard flash were fully and successfully retrieved.


\section{Data processing and  Results}
\label{sect:res}

The BBWLC acquired the linear polarization brightness of K-corona in 5500 \AA \ with passband of 50 \AA \ of the region covering altitudes from 1.08 $\text{R}_{\odot}$ to 1.50  $\text{R}_{\odot}$. The observations were planned to acquire the polarization data at angles of 0\textdegree, 45\textdegree,  90\textdegree \ and 135\textdegree \  one by one as a group. However, only polarization images of 2048$\times$2048 pixels at the first three angles were observed because of the mechanical failure of the polarizer wheel at 135\textdegree \ channel during the flight. Thus one group of data contains 30 frames of images with  exposure time of 200~ms in three directions of polarization. The total effective duration over which observations were made was approximately 1 hour and 50 minutes, in which 17,100 frames of images  were recorded. These images were first processed via dark-field subtraction and flat-field correction, then were further processed to obtain the polarized brightness ($pB$) of the corona. 

\begin{figure}[!htbp]
\centering
	\includegraphics[width=1.0\textwidth]{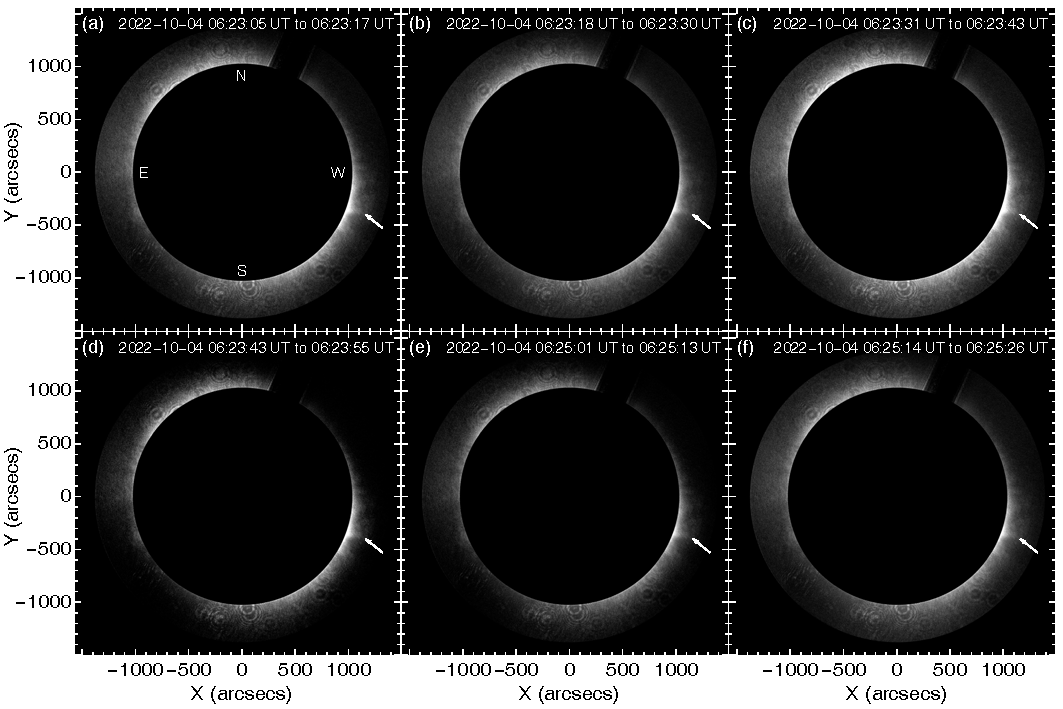}
    \caption{Time sequence of $pB$ images displaying the structure of a coronal streamer.} 
    \label{figure6}  
\end{figure}

\subsection{Polarized Brightness}
\label{sect:Pb}

Polarized light is usually described in terms of the Stokes parameters that are characterized by easy measurement and calculation since they are time-averaged values of the emission measure. We just need take care of the Stokes paremeters $I$, $Q$, and $U$ here since only the linear polarization signals of the white light corona were obtained in this work. Here,  the relationship between the signal of the detector and the parameters $(I, Q, U)$ is given as follows \citep{2004ASSL..307.....L} :

\begin{equation}
D(\beta)=(I+Qcos2\beta+Usin2\beta)/2,
\label{eq:relation}
\end{equation}
where $\beta$ is equal to  0\textdegree, 45\textdegree,  90\textdegree, respectively, and

\begin{equation}
D(0^{\circ})=(I+Q)/2,
\label{eq:D0}
\end{equation}

\begin{equation}
D(90^{\circ})=(I-Q)/2,
\label{eq:D45}
\end{equation}

\begin{equation}
D(45^{\circ})=(I+U)/2.
\label{eq:D90}
\end{equation}
Then the Stokes parameters I, Q, and U can be found: 
\begin{equation}
I=D(0^{\circ})+D(90^{\circ}),
\label{eq:I}
\end{equation}

\begin{equation}
Q=D(0^{\circ})-D(90^{\circ}),
\label{eq:Q}
\end{equation}

\begin{equation}
U=2D(45^{\circ})-D(0^{\circ})-D(90^{\circ}).
\label{eq:U}
\end{equation}

Using equations (\ref{eq:I}), (\ref{eq:Q}), and (\ref{eq:U}), we track $I$, $Q$ and $U$. Repeating this procedure for every pixel in the image, the $pB$ image can be obtained accordingly (See also \citealt{2021SoPh..296...15G})
\begin{equation}
pB=(Q^{2}+U^{2})^{1/2}.
\label{eq:pB}
\end{equation}
Meanwhile, the new normalizing-radial-graded filter proposed by \citet{2006SoPh..236..263M} was applied to enhancing images.  Finally, a time sequence of $pB$ images were obtained. Figure \ref{figure6} displays some $pB$ images for some moments around 06:23 UT. It can be seen that three main characteristics stand out on the images. First, a coronal streamer structure can be clearly seen on the southwest side of the Sun near the equator, as indicated by the white arrow in Figures \ref{figure6}(a)-\ref{figure6}(f). Second, images showed here cover the time interval of 2 min only, so no apparent change in these coronal structures could be recognized. Third, a few ring pseudo-signals could be recognized. They were caused by diffracted light from the dust on the objective lens.

We note here that a missing fourth polarization measurement may introduce extra uncertainty in the results, especially when measuring weak signals, although three linear polarization angles are mathematically sufficient to derive Stokes $I$, $Q$ and $U$. When calculating Stokes parameters $I$, $Q$ and $U$ from measurements at three polarization angles (e.g., 0\textdegree, 45\textdegree, 90\textdegree), the error inherent in the 0\textdegree \ and 90\textdegree \ measurements propagates to the derived 45\textdegree \ components. This contrasts with the scenario using four polarization angles (e.g., 0\textdegree, 45\textdegree, 90\textdegree, 135\textdegree), where the noise reduction benefits from effectively averaging over two independent pairs of measurements. Consequently, the noise level in the three-angle calculation is inherently increased by at least a factor of $\sqrt{2}$ comparing with the four-angle method. 

Meanwhile, the signal-to-noise ratio (SNR) could also estimated for our data, We define the SNR here as the ratio of the signal DN to the noise DN in the $pB$ images. The signal DN is the difference between the mean intensity within a small region containing coronal structures and the mean intensity within an area of the same size without coronal structures.  The noise DN is the standard deviation of the intensity of the same area. These two small regions should be selected at approximately the same radial distance from the solar disk center. This gives the SNR of the BBWLC of 5.6, comparing with 6.4 of that of K-Cor.

\begin{figure}[!htbp]
\centering
	\includegraphics[width=1.0\textwidth]{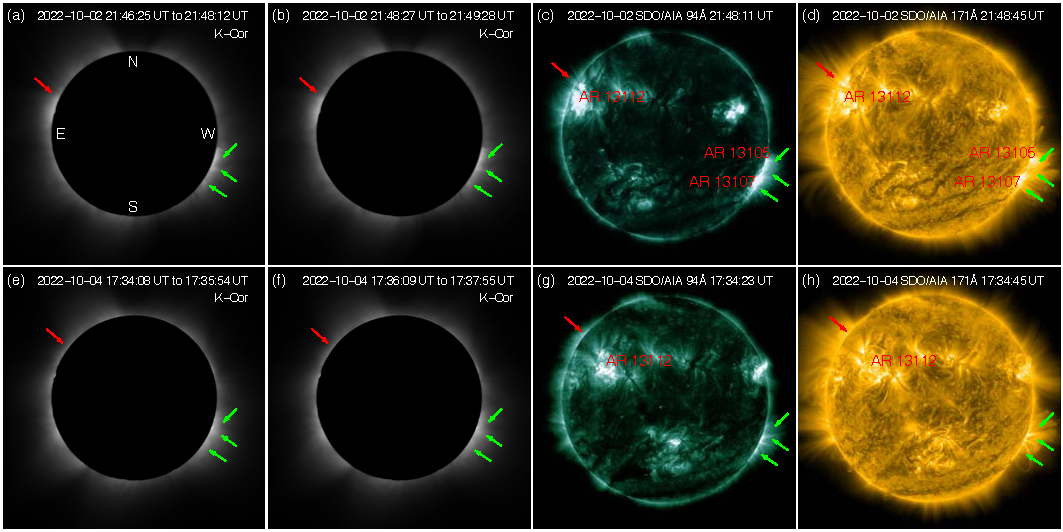}
    \caption{(a)-(b) The $pB$ images obtained by K-Cor at 21:46 UT and 21:48 UT on October 2, 2022, respectively, prior to the flight of the BBWLC. (c)-(d) Images of AIA 94 \AA \ and AIA 171 \AA \ at about the same time as Fig.\ref{figure7}(a). (e)-(f) The $pB$ images obtained by K-Cor at 17:34 UT and 17:36 UT on October 4, 2022, respectively, after the launch of BBWLC. (g)-(h) Images of AIA 94 \AA \ and AIA 171 \AA \ at about the same time as panels (e) and (f). The arrows are meant for the qualitative illustration only.} 
    \label{figure7}
\end{figure}

Due to the 18-hour time difference between Dachaidan, Qinghai and Hawaii, USA, the counterpart of our data from the K-Cor coronagraph at the Mauna Loa Solar Observatory (MLSO) in Hawaii were not available. To verify the authenticity and reliability of our data as shown in Figure \ref{figure6}, we conducted cross-checks with the K-Cor observation data that were acquired later but closest to the period of our mission. Meanwhile, we also performed cross-checks with the data obtained by the Atmospheric Imaging Assembly (AIA; \citealt{2012SoPh..275...17L}) onboard the Solar Dynamics Observatory (SDO; \citealt{2012SoPh..275....3P}) in the same period as our mission.





\subsection{Comparison of BBWLC Images with K-Cor and AIA Images}
\label{sect:Comp}

The K-Cor records the coronal $pB$ over the range from 1.05 R$_{\odot}$ to 3 $\text{R}_{\odot}$  with a spatial resolution of $11.3^{\prime\prime}$ (or $5.5^{\prime\prime}$ pixel size) and a time cadence of 15 s, while BBWLC achieves a spatial resolution of $2.8^{\prime\prime}$  (or $1.5^{\prime\prime}$ pixel size) with a time cadence of 12 s. The AIA on board SDO takes images of the Sun with an FOV of 1.3 $\text{R}_{\odot}$ in seven extreme ultraviolet channels, including 94 \AA, 131 \AA,  171 \AA,  193 \AA,  211 \AA, 304 \AA \ and 335 \AA. The 94~\AA ~and 171 \AA \ images used here are processed to level 1.5 through standard codes in the solarsoftware. These AIA images have a spatial resolution of $1.2^{\prime\prime}$ (or pixel size of $0.6^{\prime\prime}$) and a time cadence of 12 s.

Figures \ref{figure7}(a) and \ref{figure7}(b) show the $pB$ images obtained by K-Cor at 21:46 UT and 21:48 UT on October 2, 2022, respectively, prior to the flight of the BBWLC. This is the K-Cor data that we can obtain before the BBWLC launch, and it is the closest in time to the BBWLC mission period.  As can be seen, no apparent change occurred during a time interval of 2 min. Figures \ref{figure7}(c) and \ref{figure7}(d) show the images of AIA 94 \AA \ and AIA 171 \AA \ at about the same time as Figure \ref{figure7}(a). The image of the AIA indicates that two active regions were located on the northeast and southwest sides of solar disk at this time, and the structures in the $pB$ image matched to them very well (see red and green arrows).


Figures \ref{figure7}(e) and \ref{figure7}(f) show the $pB$ images obtained by K-Cor at 17:34 UT and 17:36 UT on October 4, 2022, respectively, after the launch of BBWLC. This is the K-Cor data that we can obtain after the BBWLC launch, and it is the closest in time to the BBWLC mission period. Figures \ref{figure7}(g) and \ref{figure7}(h) show the corresponding images of AIA. Comparisons of Figures \ref{figure7}(c) with \ref{figure7}(g), and \ref{figure7}(d) with \ref{figure7}(h) indicate that all the active regions already rotated by an angle due to the rotation of the Sun, and that AR 13112 moved completely onto the solar disk (see red arrow), while AR 13105 totally moved to the back of the solar disk, but part of AR 13107 was still on the limb of the Sun, leaving a small portion still visible (see green arrow). Therefore, no sign of AR 13112 showed in the $pB$ image (see red arrow), but some of the magnetic structure of AR 13107 was still recognizable (see the area indicated by the green arrows).

\begin{figure}[!htbp]
\centering
	\includegraphics[width=1.0\textwidth]{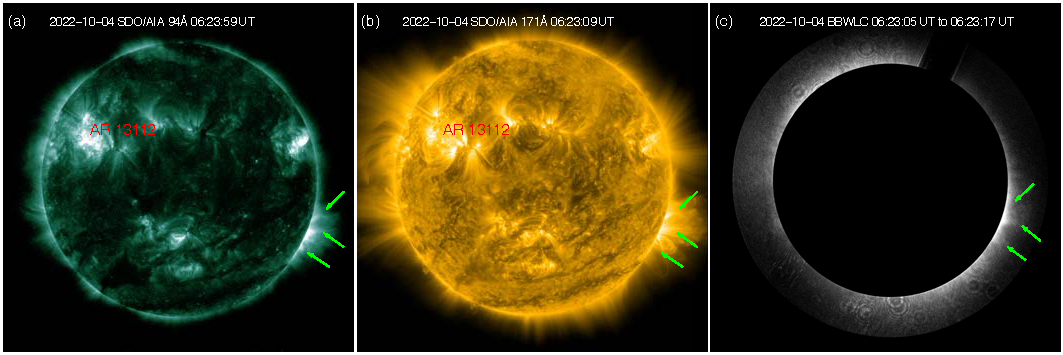}
    \caption{(a)-(b) Images of AIA 94 \AA \ and AIA 171 \AA \ at 06:23 UT on 4-Oct-2022. (c) The $pB$ images obtained by BBWLC at about the same time of (a) and (b). The arrows are meant for the qualitative illustration only.} 
    \label{figure8}  
\end{figure}

Since the observations of BBWLC (Figure \ref{figure6}) occurred before K-Cor obtained Figures \ref{figure7}(e) and \ref{figure7}(f), and AIA obtained Figures \ref{figure7}(g) and \ref{figure7}(h), most of the AR 13107 in the southwestern direction is still on the solar limb, and its extension into the corona is also recognizable by its magnetic field structure and the associated plasma distribution features (Figures \ref{figure8}(a) and \ref{figure8}(b)). For comparison, we duplicate Figure \ref{figure6}(a) in Figure \ref{figure8}(c), and notice that the magnetic structure pointed by green arrows in each panel of Figure \ref{figure8} correspond to one another very well. This confirms that the region indicated by green arrows in Figure \ref{figure8}(c) is indeed the white-light signal of the corona (i.e., the structure indicated by the white arrows in Figure \ref{figure6}), and it also corresponds well to the structure shown in the K-Cor image (Figure \ref{figure7}(e)). At the same time, it should be noted that the K-Cor, BBWLC and AIA images are not spatially co-registered, the arrows are meant for qualitative illustration only and do not indicate precise spatial correspondence across the instruments.


\section{Discussions and conclusions}
\label{sect:discussion}

With the support of “Honghu Special Fund” of the Chinese Academy of Sciences (CAS), Yunnan Observatories, together with Shandong University and  Changchun Institute of Optics, Fine Mechanics and Physics, developed a 50-mm balloon-borne white light coronagraph (BBWLC). It is used to image the K-corona in 5500 \AA \ with a passband of 50 \AA \ and a field of view (FOV) from 1.08 $\text{R}_{\odot}$ to 1.50  $\text{R}_{\odot}$ in order to observe the low coronal region that is closely related to the CME formation, corona heating and solar wind acceleration. BBWLC conducted its first science flight at Dachaidan area in Qinghai province of China on 2022 October 4, and obtained 17,100 frames of images of WLC.

By processing and analyzing these data, the K-corona signal was detected in the $p$B images around 06:23 UT on October 4, 2022, and a coronal streamer structure can be clearly seen on the southwestern side of the Sun. In order to confirm the authenticity of these signals, we compared them with the K-Cor and AIA observations, respectively, and found that these signals are in good agreement with the K-Cor and AIA observations.

We note here that this was the first experiment in our experience and we were lacking enough practice, so the temporary disruption of the communication between the headquarter and the gondola and the imperfection of the pointing mechanism resulted in that approximately 97.3\% of the images are not suitable for scientific analyses. These defects will be improved and corrected in the future experiment of BBWLC.


In addition, dust from the transportation of the equipment and from the launch site penetrated into the optical system, and despite cleaning of the objective lens prior to launch, dust still adhered to the surface of the objective lens during the launch process. Therefore, the diffracted light of dust produced some ring  pseudo-signals in the image, which reduced the signal-to-noise ratio of the observational data \citep{2006Nelson}. In the future, a number of measures will be taken to avoid the dust as  much as possible. For example, mounting the instrument in a dust-proof clean room and continuously blowing nitrogen into the lens barrels during the mounting and transportation phases so that the adhesion of dust to the individual mirrors could be reduced apparently (see also \citealt{2021SoPh..296...15G}).


We can summarize the experiments of BBWLC as follows:
(1) We have developed the first white-light coronagraph system in China, and the results of ground experiments have shown that the system meets the requirements for carrying out observations in the near-space.
(2) The white-light corona in the range from 1.08 $\text{R}_{\odot}$ to 1.50  $\text{R}_{\odot}$ was definitely observed in near space, and a streamer structure can be clearly seen in the white-light images. This is the first time that the inner corona in white-light was observed in the near-space and the corresponding data were obtained. 
(3) The experiment indicates that we possess the technology of designing and manufacturing white-light coronagraph, and have experienced observations of the white-light inner corona in the near-space.

Before we end this work, it is necessary to discuss the potential broader use of BBWLC observations in the future. Specifically, the potential of the BBWLC data for deriving physical quantities is worth elaborating. For example, one of the important physical parameters that can be derived from coronal $pB$ images is the distribution of electron density \citep{2023FrASS..1092881S}. A combination of electron density from $pB$ images with the wave phase speed derived from Doppler velocity observations from coronagraphs such as the Spectral Imaging CoronaGraph (SICG, \citealt{2025RAA....25e5013T}) and the Upgraded Coronal Multi-channel Polarimeter (UCoMP) could help derive the coronal magnetic field similar to those obtained from CoMP and UCoMP observations (e.g., see \citealt{2020Sci...369..694Y} and \citealt{2024Sci...386...76Y}). In addition, $pB$ images from coronagraphs play an important role in the study of CME structures and dynamics, as has been shown in previous work using K-Cor data (e.g., see also \citealt{2023ApJ...952L..22S} and \citealt{2023ApJ...942...19S} ). Finally, the BBWLC could also fill some missing parts of the existing instruments. For instance, the time zone offset between BBWLC’s flight operations in China and the Mauna Loa Solar Observatory (MLSO) in Hawaii enables extended monitoring of the solar corona. BBWLC and K-Cor may also offer complementary observations to each other.

\normalem
\begin{acknowledgements}

We are very grateful to the referee for the valuable comments and suggestions that helped improve this work significantly. This work was supported by the National Key R\&D Program of China No. 2022YFF0503800,  the Strategic Priority Research Program of the Chinese Academy of Sciences (grant No. XDA17040507, grant No.XDB0560000), NSFC grant 11933009, the Yunnan Province Scientist Workshop of Solar Physics, the Yunnan Key Laboratory of Solar Physics and Space Science (202205AG070009). JL benefit from the discussions of the ISSI-BJ Team “Solar eruptions: preparing for the next  generation multi-wave band coronagraphs.” Calculations in this work were carried out on the cluster in the Computational Solar Physics Laboratory of Yunnan Observatories.

\end{acknowledgements}

\bibliographystyle{raa}
\bibliography{bibtex}

@ARTICLE{2019ARA&A..57..157C,
       author = {{Cranmer}, Steven R. and {Winebarger}, Amy R.},
        title = "{The Properties of the Solar Corona and Its Connection to the Solar Wind}",
      journal = {\araa},
         year = 2019,
        month = aug,
       volume = {57},
        pages = {157-187},
          doi = {10.1146/annurev-astro-091918-104416},
archivePrefix = {arXiv},
       eprint = {1811.00461},
 primaryClass = {astro-ph.SR},
       adsurl = {https://ui.adsabs.harvard.edu/abs/2019ARA&A..57..157C}
}

@ARTICLE{1958ApJ...128..664P,
       author = {{Parker}, E.~N.},
        title = "{Dynamics of the Interplanetary Gas and Magnetic Fields.}",
      journal = {\apj},
         year = 1958,
        month = nov,
       volume = {128},
        pages = {664},
          doi = {10.1086/146579},
       adsurl = {https://ui.adsabs.harvard.edu/abs/1958ApJ...128..664P}
}

@ARTICLE{2006LRSP....3....1M,
       author = {{Marsch}, Eckart},
        title = "{Kinetic Physics of the Solar Corona and Solar Wind}",
      journal = {LRSP},
         year = 2006,
        month = dec,
       volume = {3},
       number = {1},
          eid = {1},
        pages = {1},
          doi = {10.12942/lrsp-2006-1},
       adsurl = {https://ui.adsabs.harvard.edu/abs/2006LRSP....3....1M}
}

@ARTICLE{2025A&A...695A.192C,
       author = {{Chiba}, Shota and {Shoda}, Munehito and {Imamura}, Takeshi},
        title = "{Density fluctuation in the solar corona and solar wind: A comparative analysis of radio-occultation observations and magnetohydrodynamic simulation}",
      journal = {\aap},
         year = 2025,
        month = mar,
       volume = {695},
          eid = {A192},
        pages = {A192},
          doi = {10.1051/0004-6361/202449189},
archivePrefix = {arXiv},
       eprint = {2502.11700},
 primaryClass = {astro-ph.SR},
       adsurl = {https://ui.adsabs.harvard.edu/abs/2025A&A...695A.192C}
}

@ARTICLE{2008ApJ...683.1168K,
       author = {{Ko}, Yuan-Kuen and {Li}, Jing and {Riley}, Pete and {Raymond}, John C.},
        title = "{Large-Scale Coronal Density and Abundance Structures and Their Association with Magnetic Field Structure}",
      journal = {\apj},
         year = 2008,
        month = aug,
       volume = {683},
       number = {2},
        pages = {1168-1179},
          doi = {10.1086/589873},
       adsurl = {https://ui.adsabs.harvard.edu/abs/2008ApJ...683.1168K}
}

@ARTICLE{2015A&A...583A.101M,
       author = {{Mercier}, C. and {Chambe}, G.},
        title = "{Electron density and temperature in the solar corona from multifrequency radio imaging}",
      journal = {\aap},
         year = 2015,
        month = nov,
       volume = {583},
          eid = {A101},
        pages = {A101},
          doi = {10.1051/0004-6361/201425540},
       adsurl = {https://ui.adsabs.harvard.edu/abs/2015A&A...583A.101M}
}

@ARTICLE{2016JGRA..121.7470T,
       author = {{Tomczyk}, S. and {Landi}, E. and {Burkepile}, J.~T. and {Casini}, R. and {DeLuca}, E.~E. and {Fan}, Y. and {Gibson}, S.~E. and {Lin}, H. and {McIntosh}, S.~W. and {Solomon}, S.~C. and {de Toma}, G. and {de Wijn}, A.~G. and {Zhang}, J.},
        title = "{Scientific objectives and capabilities of the Coronal Solar Magnetism Observatory}",
      journal = {JGRA},
         year = 2016,
        month = aug,
       volume = {121},
       number = {8},
        pages = {7470-7487},
          doi = {10.1002/2016JA022871},
       adsurl = {https://ui.adsabs.harvard.edu/abs/2016JGRA..121.7470T}
}

@ARTICLE{2001ApJ...548.1081H,
       author = {{Hayes}, A.~P. and {Vourlidas}, A. and {Howard}, R.~A.},
        title = "{Deriving the Electron Density of the Solar Corona from the Inversion of Total Brightness Measurements}",
      journal = {\apj},
         year = 2001,
        month = feb,
       volume = {548},
       number = {2},
        pages = {1081-1086},
          doi = {10.1086/319029},
       adsurl = {https://ui.adsabs.harvard.edu/abs/2001ApJ...548.1081H}
}

@ARTICLE{2002ApJ...576..485L,
       author = {{Lin}, J. and {van Ballegooijen}, A.~A.},
        title = "{Catastrophic and Noncatastrophic Mechanisms for Coronal Mass Ejections}",
      journal = {\apj},
         year = 2002,
        month = sep,
       volume = {576},
       number = {1},
        pages = {485-492},
          doi = {10.1086/341737},
       adsurl = {https://ui.adsabs.harvard.edu/abs/2002ApJ...576..485L}
}

@ARTICLE{2003NewAR..47...53L,
       author = {{Lin}, J. and {Soon}, W. and {Baliunas}, S.~L.},
        title = "{Theories of solar eruptions: a review}",
      journal = {\nar},
         year = 2003,
        month = apr,
       volume = {47},
       number = {2},
        pages = {53-84},
          doi = {10.1016/S1387-6473(02)00271-3},
       adsurl = {https://ui.adsabs.harvard.edu/abs/2003NewAR..47...53L}
}

@ARTICLE{2023SSPMA..53y9611J,
       author = {{Lin}, J. and {Song}, T.~F. and {Sun}, M.~Z. and {Zhang}, T.},
        title = "{A 50-mm balloon-borne white-light coronagraph: I.Basic structure and experiments on the ground}",
      journal = {Sci. Sin. Phys. Mech. Astron.},
         year = 2023,
        month = may,
       volume = {53},
       number = {5},
        pages = {259611},
          doi = {10.1360/SSPMA-2022-0363},
       adsurl = {https://ui.adsabs.harvard.edu/abs/2023SSPMA..53y9611J}
}

@ARTICLE{2006SSRv..123..251F,
       author = {{Forbes}, T.~G. and {Linker}, J.~A. and {Chen}, J. and {Cid}, C. and {K{\'o}ta}, J. and {Lee}, M.~A. and {Mann}, G. and {Miki{\'c}}, Z. and {Potgieter}, M.~S. and {Schmidt}, J.~M. and {Siscoe}, G.~L. and {Vainio}, R. and {Antiochos}, S.~K. and {Riley}, P.},
        title = "{CME Theory and Models}",
      journal = {\ssr},
         year = 2006,
        month = mar,
       volume = {123},
       number = {1-3},
        pages = {251-302},
          doi = {10.1007/s11214-006-9019-8},
       adsurl = {https://ui.adsabs.harvard.edu/abs/2006SSRv..123..251F}
}

@ARTICLE{2011LRSP....8....1C,
       author = {{Chen}, P.~F.},
        title = "{Coronal Mass Ejections: Models and Their Observational Basis}",
      journal = {LRSP},
         year = 2011,
        month = dec,
       volume = {8},
       number = {1},
          eid = {1},
        pages = {1},
          doi = {10.12942/lrsp-2011-1},
       adsurl = {https://ui.adsabs.harvard.edu/abs/2011LRSP....8....1C}
}

@ARTICLE{2013SoPh..288..637T,
       author = {{Tian}, H. and {Tomczyk}, S. and {McIntosh}, S.~W. and {Bethge}, C. and {de Toma}, G. and {Gibson}, S.},
        title = "{Observations of Coronal Mass Ejections with the Coronal Multichannel Polarimeter}",
      journal = {\solphys},
         year = 2013,
        month = dec,
       volume = {288},
       number = {2},
        pages = {637-650},
          doi = {10.1007/s11207-013-0317-5},
archivePrefix = {arXiv},
       eprint = {1303.4647},
 primaryClass = {astro-ph.SR},
       adsurl = {https://ui.adsabs.harvard.edu/abs/2013SoPh..288..637T}
}

@ARTICLE{1976SoPh...48..389G,
       author = {{Gosling}, J.~T. and {Hildner}, E. and {MacQueen}, R.~M. and {Munro}, R.~H. and {Poland}, A.~I. and {Ross}, C.~L.},
        title = "{The speeds of coronal mass ejection events.}",
      journal = {\solphys},
         year = 1976,
        month = jun,
       volume = {48},
       number = {2},
        pages = {389-397},
          doi = {10.1007/BF00152004},
       adsurl = {https://ui.adsabs.harvard.edu/abs/1976SoPh...48..389G}
}

@ARTICLE{1993JGR....9813177H,
       author = {{Hundhausen}, A.~J.},
        title = "{Sizes and locations of coronal mass ejections: SMM observations from 1980 and 1984-1989}",
      journal = {JGR},
         year = 1993,
        month = aug,
       volume = {98},
       number = {A8},
        pages = {13177-13200},
          doi = {10.1029/93JA00157},
       adsurl = {https://ui.adsabs.harvard.edu/abs/1993JGR....9813177H}
}

@ARTICLE{1999JGR...10412493S,
       author = {{St. Cyr}, O.~C. and {Burkepile}, J.~T. and {Hundhausen}, A.~J. and {Lecinski}, A.~R.},
        title = "{A comparison of ground-based and spacecraft observations of coronal mass ejections from 1980-1989}",
      journal = {JGR},
         year = 1999,
        month = jun,
       volume = {104},
       number = {A6},
        pages = {12493-12506},
          doi = {10.1029/1999JA900045},
       adsurl = {https://ui.adsabs.harvard.edu/abs/1999JGR...10412493S}
}

@INPROCEEDINGS{2012SPIE.8444E..3ND,
       author = {{de Wijn}, Alfred G. and {Burkepile}, Joan T. and {Tomczyk}, Steven and {Nelson}, Peter G. and {Huang}, Pei and {Gallagher}, Dennis},
        title = "{Stray light and polarimetry considerations for the COSMO K-Coronagraph}",
    booktitle = {Ground-based and Airborne Telescopes IV},
         year = 2012,
       editor = {{Stepp}, Larry M. and {Gilmozzi}, Roberto and {Hall}, Helen J.},
       series = {Society of Photo-Optical Instrumentation Engineers (SPIE) Conference Series},
       volume = {8444},
        month = sep,
          eid = {84443N},
        pages = {84443N},
          doi = {10.1117/12.926511},
archivePrefix = {arXiv},
       eprint = {1207.0978},
 primaryClass = {astro-ph.IM},
       adsurl = {https://ui.adsabs.harvard.edu/abs/2012SPIE.8444E..3ND}
}

@ARTICLE{1939MNRAS..99..580L,
       author = {{Lyot}, Bernard},
        title = "{The study of the solar corona and prominences without eclipses (George Darwin Lecture, 1939)}",
      journal = {\mnras},
         year = 1939,
        month = jun,
       volume = {99},
        pages = {580},
          doi = {10.1093/mnras/99.8.580},
       adsurl = {https://ui.adsabs.harvard.edu/abs/1939MNRAS..99..580L}
}

@ARTICLE{1998EP&S...50..493K,
       author = {{Kimura}, Hiroshi and {Mann}, Ingrid},
        title = "{Brightness of the solar F-corona}",
      journal = {Earth, Planets and Space},
         year = 1998,
        month = jun,
       volume = {50},
       number = {6-7},
        pages = {493-499},
          doi = {10.1186/BF03352140},
       adsurl = {https://ui.adsabs.harvard.edu/abs/1998EP&S...50..493K}
}

@ARTICLE{2008JASTP..70..356S,
       author = {{Shopov}, Y.~Y. and {Stoykova}, D.~A. and {Stoitchkova}, K. and {Tsankov}, L.~T. and {Tanev}, A. and {Burin}, Kl and {Belchev}, St and {Rusanov}, V. and {Ivanov}, D. and {Stoev}, A. and {Muglova}, P. and {Iliev}, I.},
        title = "{Structure of the solar dust corona and its interaction with the other coronal components}",
      journal = {JASTP},
         year = 2008,
        month = feb,
       volume = {70},
       number = {2-4},
        pages = {356-364},
          doi = {10.1016/j.jastp.2007.08.058},
archivePrefix = {arXiv},
       eprint = {0909.1722},
 primaryClass = {astro-ph.SR},
       adsurl = {https://ui.adsabs.harvard.edu/abs/2008JASTP..70..356S}
}

@ARTICLE{1999PASJ...51..269S,
       author = {{Singh}, Jagdev and {Ichimoto}, Kiyoshi and {Imai}, Hideki and {Sakurai}, Takashi and {Takeda}, Aki},
        title = "{Spectroscopic Studies of the Solar Corona I. Spatial Variations in Line Parameters of Green and Red Coronal Lines}",
      journal = {\pasj},
         year = 1999,
        month = apr,
       volume = {51},
        pages = {269-276},
          doi = {10.1093/pasj/51.2.269},
       adsurl = {https://ui.adsabs.harvard.edu/abs/1999PASJ...51..269S}
}

@ARTICLE{2019Natur.576..232H,
       author = {{Howard}, R.~A. and {Vourlidas}, A. and {Bothmer}, V. and {Colaninno}, R.~C. and {DeForest}, C.~E. and {Gallagher}, B. and {Hall}, J.~R. and {Hess}, P. and {Higginson}, A.~K. and {Korendyke}, C.~M. and {Kouloumvakos}, A. and {Lamy}, P.~L. and {Liewer}, P.~C. and {Linker}, J. and {Linton}, M. and {Penteado}, P. and {Plunkett}, S.~P. and {Poirier}, N. and {Raouafi}, N.~E. and {Rich}, N. and {Rochus}, P. and {Rouillard}, A.~P. and {Socker}, D.~G. and {Stenborg}, G. and {Thernisien}, A.~F. and {Viall}, N.~M.},
        title = "{Near-Sun observations of an F-corona decrease and K-corona fine structure}",
      journal = {\nat},
         year = 2019,
        month = dec,
       volume = {576},
       number = {7786},
        pages = {232-236},
          doi = {10.1038/s41586-019-1807-x},
       adsurl = {https://ui.adsabs.harvard.edu/abs/2019Natur.576..232H}
}

@ARTICLE{1975ApOpt..14..743K,
       author = {{Koomen}, M.~J. and {Detwiler}, C.~R. and {Brueckner}, G.~E. and {Cooper}, H.~W. and {Tousey}, R.},
        title = "{White light coronagraph in OSO - 7}",
      journal = {\ao},
         year = 1975,
        month = mar,
       volume = {14},
       number = {3},
        pages = {743-751},
          doi = {10.1364/AO.14.000743},
       adsurl = {https://ui.adsabs.harvard.edu/abs/1975ApOpt..14..743K}
}

@ARTICLE{1981ApJ...244L.117H,
       author = {{House}, L.~L. and {Wagner}, W.~J. and {Hildner}, E. and {Sawyer}, C. and {Schmidt}, H.~U.},
        title = "{Studies of the corona with the Solar Maximum Mission coronagraph/polarimeter}",
      journal = {\apjl},
         year = 1981,
        month = mar,
       volume = {244},
        pages = {L117-L121},
          doi = {10.1086/183494},
       adsurl = {https://ui.adsabs.harvard.edu/abs/1981ApJ...244L.117H}
}

@ARTICLE{1974JGR....79.4581G,
       author = {{Gosling}, J.~T. and {Hildner}, E. and {MacQueen}, R.~M. and {Munro}, R.~H. and {Poland}, A.~I. and {Ross}, C.~L.},
        title = "{Mass ejections from the Sun: A view from Skylab}",
      journal = {JGR},
         year = 1974,
        month = nov,
       volume = {79},
       number = {31},
        pages = {4581},
          doi = {10.1029/JA079i031p04581},
       adsurl = {https://ui.adsabs.harvard.edu/abs/1974JGR....79.4581G}
}

@ARTICLE{1995SoPh..162..357B,
       author = {{Brueckner}, G.~E. and {Howard}, R.~A. and {Koomen}, M.~J. and {Korendyke}, C.~M. and {Michels}, D.~J. and {Moses}, J.~D. and {Socker}, D.~G. and {Dere}, K.~P. and {Lamy}, P.~L. and {Llebaria}, A. and {Bout}, M.~V. and {Schwenn}, R. and {Simnett}, G.~M. and {Bedford}, D.~K. and {Eyles}, C.~J.},
        title = "{The Large Angle Spectroscopic Coronagraph (LASCO)}",
      journal = {\solphys},
         year = 1995,
        month = dec,
       volume = {162},
       number = {1-2},
        pages = {357-402},
          doi = {10.1007/BF00733434},
       adsurl = {https://ui.adsabs.harvard.edu/abs/1995SoPh..162..357B}
}

@ARTICLE{2004JGRA..109.7105Y,
       author = {{Yashiro}, S. and {Gopalswamy}, N. and {Michalek}, G. and {St. Cyr}, O.~C. and {Plunkett}, S.~P. and {Rich}, N.~B. and {Howard}, R.~A.},
        title = "{A catalog of white light coronal mass ejections observed by the SOHO spacecraft}",
      journal = {JGRA},
         year = 2004,
        month = jul,
       volume = {109},
       number = {A7},
          eid = {A07105},
        pages = {A07105},
          doi = {10.1029/2003JA010282},
       adsurl = {https://ui.adsabs.harvard.edu/abs/2004JGRA..109.7105Y}
}

@INPROCEEDINGS{2003SPIE.4853....1T,
       author = {{Thompson}, William T. and {Davila}, Joseph M. and {Fisher}, Richard R. and {Orwig}, Larry E. and {Mentzell}, John E. and {Hetherington}, Samuel E. and {Derro}, Rebecca J. and {Federline}, Robert E. and {Clark}, David C. and {Chen}, Philip T.~C. and {Tveekrem}, June L. and {Martino}, Anthony J. and {Novello}, Joseph and {Wesenberg}, Richard P. and {StCyr}, Orville C. and {Reginald}, Nelson L. and {Howard}, Russell A. and {Mehalick}, Kimberly I. and {Hersh}, Michael J. and {Newman}, Miles D. and {Thomas}, Debbie L. and {Card}, Gregory L. and {Elmore}, David F.},
        title = "{COR1 inner coronagraph for STEREO-SECCHI}",
    booktitle = {Innovative Telescopes and Instrumentation for Solar Astrophysics},
         year = 2003,
       editor = {{Keil}, Stephen L. and {Avakyan}, Sergey V.},
       series = {Society of Photo-Optical Instrumentation Engineers (SPIE) Conference Series},
       volume = {4853},
        month = feb,
        pages = {1-11},
          doi = {10.1117/12.460267},
       adsurl = {https://ui.adsabs.harvard.edu/abs/2003SPIE.4853....1T}
}

@ARTICLE{2011SoPh..268....1B,
       author = {{Barthol}, P. and {Gandorfer}, A. and {Solanki}, S.~K. and {Sch{\"u}ssler}, M. and {Chares}, B. and {Curdt}, W. and {Deutsch}, W. and {Feller}, A. and {Germerott}, D. and {Grauf}, B. and {Heerlein}, K. and {Hirzberger}, J. and {Kolleck}, M. and {Meller}, R. and {M{\"u}ller}, R. and {Riethm{\"u}ller}, T.~L. and {Tomasch}, G. and {Kn{\"o}lker}, M. and {Lites}, B.~W. and {Card}, G. and {Elmore}, D. and {Fox}, J. and {Lecinski}, A. and {Nelson}, P. and {Summers}, R. and {Watt}, A. and {Mart{\'\i}nez Pillet}, V. and {Bonet}, J.~A. and {Schmidt}, W. and {Berkefeld}, T. and {Title}, A.~M. and {Domingo}, V. and {Gasent Blesa}, J.~L. and {del Toro Iniesta}, J.~C. and {L{\'o}pez Jim{\'e}nez}, A. and {{\'A}lvarez-Herrero}, A. and {Sabau-Graziati}, L. and {Widani}, C. and {Haberler}, P. and {H{\"a}rtel}, K. and {Kampf}, D. and {Levin}, T. and {P{\'e}rez Grande}, I. and {Sanz-Andr{\'e}s}, A. and {Schmidt}, E.},
        title = "{The Sunrise Mission}",
      journal = {\solphys},
         year = 2011,
        month = jan,
       volume = {268},
       number = {1},
        pages = {1-34},
          doi = {10.1007/s11207-010-9662-9},
archivePrefix = {arXiv},
       eprint = {1009.2689},
 primaryClass = {astro-ph.IM},
       adsurl = {https://ui.adsabs.harvard.edu/abs/2011SoPh..268....1B}
}

@ARTICLE{2024ChJSS..44.1068L,
       author = {{Li}, Yijian and {Huang}, Wanning and {Zhou}, Jianghua and {Zhang}, Xiaojun and {Zhang}, Hangyue},
        title = "{Development Status and Prospects of Near Space Observatories}",
      journal = {CJSS},
         year = 2024,
        month = nov,
       volume = {44},
       number = {6},
        pages = {1068-1085},
          doi = {10.11728/cjss2024.06.2023-0145},
       adsurl = {https://ui.adsabs.harvard.edu/abs/2024ChJSS..44.1068L}
}

@ARTICLE{1959ApJ...130..345S,
       author = {{Schwarzschild}, Martin},
        title = "{Photographs of the Solar Granulation Taken from the Stratosphere.}",
      journal = {\apj},
         year = 1959,
        month = sep,
       volume = {130},
        pages = {345},
          doi = {10.1086/146725},
       adsurl = {https://ui.adsabs.harvard.edu/abs/1959ApJ...130..345S}
}

@ARTICLE{1972SoPh...26..305K,
       author = {{Krat}, V.~A. and {Karpinsky}, V.~N. and {Pravdjuk}, L.~M.},
        title = "{On the Sunspot Structure}",
      journal = {\solphys},
         year = 1972,
        month = oct,
       volume = {26},
       number = {2},
        pages = {305-317},
          doi = {10.1007/BF00165272},
       adsurl = {https://ui.adsabs.harvard.edu/abs/1972SoPh...26..305K}
}

@ARTICLE{1978PASJ...30..337H,
       author = {{Hirayama}, T.},
        title = "{A Model of Solar Faculae and Their Lifetime}",
      journal = {\pasj},
         year = 1978,
        month = jan,
       volume = {30},
        pages = {337-352},
       adsurl = {https://ui.adsabs.harvard.edu/abs/1978PASJ...30..337H}
}

@ARTICLE{1979SoPh...63...35H,
       author = {{Herse}, M.},
        title = "{High resolution photographs of the sun near 200 nm.}",
      journal = {\solphys},
         year = 1979,
        month = aug,
       volume = {63},
       number = {1},
        pages = {35-60},
          doi = {10.1007/BF00155693},
       adsurl = {https://ui.adsabs.harvard.edu/abs/1979SoPh...63...35H}
}

@ARTICLE{1996SoPh..164..403R,
       author = {{Rust}, D.~M. and {Murphy}, G. and {Strohbehn}, K. and {Keller}, C.~U.},
        title = "{Balloon-Borne Polarimetry}",
      journal = {\solphys},
         year = 1996,
        month = mar,
       volume = {164},
       number = {1-2},
        pages = {403-415},
          doi = {10.1007/BF00146652},
       adsurl = {https://ui.adsabs.harvard.edu/abs/1996SoPh..164..403R}
}

@ARTICLE{2004AdSpR..33.1746B,
       author = {{Bernasconi}, P.~N. and {Eaton}, H.~A.~C. and {Foukal}, P. and {Rust}, D.~M.},
        title = "{The solar bolometric imager}",
      journal = {AdSpR},
         year = 2004,
        month = jan,
       volume = {33},
       number = {10},
        pages = {1746-1754},
          doi = {10.1016/j.asr.2003.08.028},
       adsurl = {https://ui.adsabs.harvard.edu/abs/2004AdSpR..33.1746B}
}

@ARTICLE{2021SoPh..296...15G,
       author = {{Gopalswamy}, N. and {Newmark}, J. and {Yashiro}, S. and {M{\"a}kel{\"a}}, P. and {Reginald}, N. and {Thakur}, N. and {Gong}, Q. and {Kim}, Y. -H. and {Cho}, K. -S. and {Choi}, S. -H. and {Baek}, J. -H. and {Bong}, S. -C. and {Yang}, H. -S. and {Park}, J. -Y. and {Kim}, J. -H. and {Park}, Y. -D. and {Lee}, J. -O. and {Kim}, R. -S. and {Lim}, E. -K.},
        title = "{The Balloon-Borne Investigation of Temperature and Speed of Electrons in the Corona (BITSE): Mission Description and Preliminary Results}",
      journal = {\solphys},
         year = 2021,
        month = jan,
       volume = {296},
       number = {1},
          eid = {15},
        pages = {15},
          doi = {10.1007/s11207-020-01751-8},
archivePrefix = {arXiv},
       eprint = {2011.06111},
 primaryClass = {astro-ph.SR},
       adsurl = {https://ui.adsabs.harvard.edu/abs/2021SoPh..296...15G}
}

@ARTICLE{2018SoPh..293....1Z,
       author = {{Zhao}, M.~Y. and {Liu}, Y. and {Elmhamdi}, A. and {Kordi}, A.~S. and {Zhang}, X.~F. and {Song}, T.~F. and {Tian}, Z.~J.},
        title = "{Conditions for Coronal Observations at the Lijiang Observatory in 2011}",
      journal = {\solphys},
         year = 2018,
        month = jan,
       volume = {293},
       number = {1},
          eid = {1},
        pages = {1},
          doi = {10.1007/s11207-017-1223-z},
archivePrefix = {arXiv},
       eprint = {1709.02053},
 primaryClass = {astro-ph.IM},
       adsurl = {https://ui.adsabs.harvard.edu/abs/2018SoPh..293....1Z}
}

@INPROCEEDINGS{2018SPIE10704E..22L,
       author = {{Liu}, Yu and {Li}, Xiaobo and {Zhang}, Xuefei and {Song}, Tengfei and {Wang}, Jingxing and {Zhao}, Mingyu and {Xia}, Lidong and {Song}, Qiwu},
        title = "{Operation of the astronomical monitoring stations at Mt. Wumingshan}",
    booktitle = {Observatory Operations: Strategies, Processes, and Systems VII},
         year = 2018,
       series = {Society of Photo-Optical Instrumentation Engineers (SPIE) Conference Series},
       volume = {10704},
        month = jul,
          eid = {1070422},
        pages = {1070422},
          doi = {10.1117/12.2309831},
       adsurl = {https://ui.adsabs.harvard.edu/abs/2018SPIE10704E..22L}
}

@ARTICLE{2020RAA....20...85S,
       author = {{Song}, T. F. and {Liu}, Yu and {Wang}, J. X. and {Zhang}, Xue-Fei and {Liu}, Shun-Qing and {Zhao}, Ming-Yu and {Li}, Xiao-Bo and {Cai}, Zhan-Chuan and {Song}, Qi-Wu and {Cao}, Zi-Huang and {Ruan}, Yu},
        title = "{Site testing campaign for the Large Optical/infrared Telescope of China: general introduction of the Daocheng site}",
      journal = {RAA},
         year = 2020,
        month = jun,
       volume = {20},
       number = {6},
          eid = {085},
        pages = {085},
          doi = {10.1088/1674-4527/20/6/85},
       adsurl = {https://ui.adsabs.harvard.edu/abs/2020RAA....20...85S}
}

@ARTICLE{2021MNRAS.505.3070S,
       author = {{Song}, T. F. and {Liu}, Yu and {Cai}, Z. C. and {Zhao}, Mingyu and {Zhang}, Xuefei and {Wang}, Jingxing and {Li}, Xiaobo and {Huang}, Shanjie and {Song}, Qiwu and {Du}, Zhimao},
        title = "{Evaluation of the day-time ground-level turbulence at Mt Wumingshan with a microthermal sensor}",
      journal = {\mnras},
         year = 2021,
        month = aug,
       volume = {505},
       number = {2},
        pages = {3070-3077},
          doi = {10.1093/mnras/stab1469},
       adsurl = {https://ui.adsabs.harvard.edu/abs/2021MNRAS.505.3070S}
}

@ARTICLE{2006Nelson,
       author = {{Nelson}, P.~G.},
        title = "{An analysis of scattered light in reflecting and refracting primary objectives for coronagraphs}",
      journal = {COSMO Technical Note 8},
         year = 2006,
          doi = {https://opensky.ucar.edu/system/files/2024-09/reports_12.pdf},
       adsurl = {https://ui.adsabs.harvard.edu/abs/2021MNRAS.505.3070S}
}

@ARTICLE{2024RAA....24b5020L,
       author = {{Liu}, D. Y. and {Zhang}, H. X. and {Sun}, M. Z and {Huang}, Zheng-Hua and {Xia}, Li-Dong and {Liu}, Wei-Xin and {Fu}, Hui},
        title = "{A New Method for Monitoring Scattered Stray Light of an Inner-occulted Coronagraph}",
      journal = {RAA},
         year = 2024,
        month = feb,
       volume = {24},
       number = {2},
          eid = {025020},
        pages = {025020},
          doi = {10.1088/1674-4527/ad019c},
       adsurl = {https://ui.adsabs.harvard.edu/abs/2024RAA....24b5020L}
}

@ARTICLE{1985OptEn..24..380B,
       author = {{Bennett}, Jean M.},
        title = "{Comparison Of Techniques For Measuring The Roughness Of Optical Surfaces}",
      journal =  {OptEn},
         year = 1985,
        month = jun,
       volume = {24},
       number = {3},
        pages = {380},
          doi = {10.1117/12.7973493},
       adsurl = {https://ui.adsabs.harvard.edu/abs/1985OptEn..24..380B}
}

@BOOK{2004ASSL..307.....L,
       author = {{Landi Degl'Innocenti}, E. and {Landolfi}, M.},
        title = "{Polarization in Spectral Lines}",
         year = {2004},
       volume = {307},
        pages = {19},
          doi = {10.1007/978-1-4020-2415-3},
       adsurl = {https://ui.adsabs.harvard.edu/abs/2004ASSL..307.....L}
}

@ARTICLE{2006SoPh..236..263M,
       author = {{Morgan}, Huw and {Habbal}, Shadia Rifai and {Woo}, Richard},
        title = "{The Depiction of Coronal Structure in White-Light Images}",
      journal = {\solphys},
         year = 2006,
        month = jul,
       volume = {236},
       number = {2},
        pages = {263-272},
          doi = {10.1007/s11207-006-0113-6},
archivePrefix = {arXiv},
       eprint = {astro-ph/0602174},
 primaryClass = {astro-ph},
       adsurl = {https://ui.adsabs.harvard.edu/abs/2006SoPh..236..263M}
}

@ARTICLE{2012SoPh..275...17L,
       author = {{Lemen}, James R. and {Title}, Alan M. and {Akin}, David J. and {Boerner}, Paul F. and {Chou}, Catherine and {Drake}, Jerry F. and {Duncan}, Dexter W. and {Edwards}, Christopher G. and {Friedlaender}, Frank M. and {Heyman}, Gary F. and {Hurlburt}, Neal E. and {Katz}, Noah L. and {Kushner}, Gary D. and {Levay}, Michael and {Lindgren}, Russell W. and {Mathur}, Dnyanesh P. and {McFeaters}, Edward L. and {Mitchell}, Sarah and {Rehse}, Roger A. and {Schrijver}, Carolus J. and {Springer}, Larry A. and {Stern}, Robert A. and {Tarbell}, Theodore D. and {Wuelser}, Jean-Pierre and {Wolfson}, C. Jacob and {Yanari}, Carl and {Bookbinder}, Jay A. and {Cheimets}, Peter N. and {Caldwell}, David and {Deluca}, Edward E. and {Gates}, Richard and {Golub}, Leon and {Park}, Sang and {Podgorski}, William A. and {Bush}, Rock I. and {Scherrer}, Philip H. and {Gummin}, Mark A. and {Smith}, Peter and {Auker}, Gary and {Jerram}, Paul and {Pool}, Peter and {Soufli}, Regina and {Windt}, David L. and {Beardsley}, Sarah and {Clapp}, Matthew and {Lang}, James and {Waltham}, Nicholas},
        title = "{The Atmospheric Imaging Assembly (AIA) on the Solar Dynamics Observatory (SDO)}",
      journal = {\solphys},
         year = 2012,
        month = jan,
       volume = {275},
       number = {1-2},
        pages = {17-40},
          doi = {10.1007/s11207-011-9776-8},
       adsurl = {https://ui.adsabs.harvard.edu/abs/2012SoPh..275...17L}
}

@ARTICLE{2012SoPh..275....3P,
       author = {{Pesnell}, W. Dean and {Thompson}, B.~J. and {Chamberlin}, P.~C.},
        title = "{The Solar Dynamics Observatory (SDO)}",
      journal = {\solphys},
         year = 2012,
        month = jan,
       volume = {275},
       number = {1-2},
        pages = {3-15},
          doi = {10.1007/s11207-011-9841-3},
       adsurl = {https://ui.adsabs.harvard.edu/abs/2012SoPh..275....3P}
}

@INPROCEEDINGS{2021SPIE12070E..0BL,
       author = {{Liu}, Yu and {Zhang}, Xuefei and {Song}, Tengfei and {Sun}, Mingzhe and {Liu}, Dayang and {Wang}, Jingxing and {Zhao}, Mingyu and {Zhang}, Tao and {Xu}, Fangyu and {Fu}, Honglin and {Pi}, Xiaoyu and {Huang}, Shanjie and {Li}, Yan and {Fu}, Yu and {Fan}, Jiankang and {Liu}, Shunqing and {Shen}, Yuandeng and {Sha}, Feiyang and {Li}, Yuqiang and {Jin}, Zhenyu and {Liu}, Zhong and {Xia}, Lidong and {Zhang}, Hongxin and {Huang}, Min and {Liu}, Yang and {Wang}, Min and {Li}, Shaokun and {Lin}, Jun},
        title = "{Ground experiment of a 50 mm balloon-borne coronagraph for near space project}",
    booktitle = {10th International Symposium on Advanced Optical Manufacturing and Testing Technologies: Large Mirror and Telescopes},
         year = 2021,
       editor = {{Rao}, Chang-Hui and {Veillet}, Christian and {Ma}, Xiaoliang and {Fan}, Bin and {Liu}, Fengchuan and {Collados Vera}, Manuel},
       series = {Society of Photo-Optical Instrumentation Engineers (SPIE) Conference Series},
       volume = {12070},
        month = dec,
          eid = {120700B},
        pages = {120700B},
          doi = {10.1117/12.2605310},
       adsurl = {https://ui.adsabs.harvard.edu/abs/2021SPIE12070E..0BL}
}

@article{Liu2025,
doi = {10.1088/1674-4527/ad9a34},
url = {https://dx.doi.org/10.1088/1674-4527/ad9a34},
year = {2025},
month = {jan},
publisher = {National Astromonical Observatories, CAS and IOP Publishing},
volume = {25},
number = {1},
pages = {015014},
author = {Liu, Da Yang and Yu, Xiao Yu and Zhang, Hong Xin and Huang, Zheng-Hua and Xia, Li-Dong and Sun, Ming-Zhe and Mao, Xian-Liang and Sun, Bo-Yu and Tang, Ning and Fu, Hui and Liu, Wei-Xin and Zhang, Chao and Han, Jian-Ping},
title = {Study on Real-time Monitoring Method for Dust-scattered Stray Light in the Spectral Imaging CoronaGraph of the Chinese Meridian Project Phase II},
journal = {RAA}
}

@ARTICLE{2023ApJ...942...19S,
       author = {{Song}, Hongqiang and {Zhang}, Jie and {Li}, Leping and {Yang}, Zihao and {Xia}, Lidong and {Zheng}, Ruisheng and {Chen}, Yao},
        title = "{On the Nature of the Three-part Structure of Solar Coronal Mass Ejections}",
      journal = {\apj},
         year = 2023,
        month = jan,
       volume = {942},
       number = {1},
          eid = {19},
        pages = {19},
          doi = {10.3847/1538-4357/aca6e0},
archivePrefix = {arXiv},
       eprint = {2212.04013},
 primaryClass = {astro-ph.SR},
       adsurl = {https://ui.adsabs.harvard.edu/abs/2023ApJ...942...19S}
}

@ARTICLE{2023ApJ...952L..22S,
       author = {{Song}, Hongqiang and {Li}, Leping and {Zhou}, Zhenjun and {Xia}, Lidong and {Cheng}, Xin and {Chen}, Yao},
        title = "{The Structure of Coronal Mass Ejections Recorded by the K-Coronagraph at Mauna Loa Solar Observatory}",
      journal = {\apjl},
         year = 2023,
        month = jul,
       volume = {952},
       number = {1},
          eid = {L22},
        pages = {L22},
          doi = {10.3847/2041-8213/ace422},
archivePrefix = {arXiv},
       eprint = {2307.01398},
 primaryClass = {astro-ph.SR},
       adsurl = {https://ui.adsabs.harvard.edu/abs/2023ApJ...952L..22S}
}

@ARTICLE{2023FrASS..1092881S,
       author = {{Sheoran}, Jyoti and {Pant}, Vaibhav and {Patel}, Ritesh and {Banerjee}, Dipankar},
        title = "{Evolution of the Thermodynamic Properties of a Coronal Mass Ejection in the Inner Corona}",
      journal = {FrASS},
         year = 2023,
        month = feb,
       volume = {10},
          eid = {27},
        pages = {27},
          doi = {10.3389/fspas.2023.1092881},
archivePrefix = {arXiv},
       eprint = {2301.13184},
 primaryClass = {astro-ph.SR},
       adsurl = {https://ui.adsabs.harvard.edu/abs/2023FrASS..1092881S}
}

@ARTICLE{2025RAA....25e5013T,
       author = {{Tang}, Ning and {Liu}, Wei-Xin and {Mao}, Xian-Liang and {Yu}, Xiao-Yu and {Zuo}, Xiu-Hui and {Sun}, Bo-Yu and {Zheng}, Hao-Ran and {Liu}, Da-Yang and {Zhang}, Xue-Fei and {Sun}, Ming-Zhe and {Huang}, Zheng-Hua and {Fu}, Hui and {Xia}, Li-Dong},
        title = "{Photometric Calibration of Spectral Imaging CoronaGraph with Dual-wavelength Observation}",
      journal = {RAA},
         year = 2025,
        month = may,
       volume = {25},
       number = {5},
          eid = {055013},
        pages = {055013},
          doi = {10.1088/1674-4527/adcc7d},
       adsurl = {https://ui.adsabs.harvard.edu/abs/2025RAA....25e5013T}
}

@ARTICLE{2024Sci...386...76Y,
       author = {{Yang}, Zihao and {Tian}, Hui and {Tomczyk}, Steven and {Liu}, Xianyu and {Gibson}, Sarah and {Morton}, Richard J. and {Downs}, Cooper},
        title = "{Observing the evolution of the Sun's global coronal magnetic field over 8 months}",
      journal = {Science},
         year = 2024,
        month = oct,
       volume = {386},
       number = {6717},
        pages = {76-82},
          doi = {10.1126/science.ado2993},
archivePrefix = {arXiv},
       eprint = {2410.16555},
 primaryClass = {astro-ph.SR},
       adsurl = {https://ui.adsabs.harvard.edu/abs/2024Sci...386...76Y}
}

@ARTICLE{2020Sci...369..694Y,
       author = {{Yang}, Zihao and {Bethge}, Christian and {Tian}, Hui and {Tomczyk}, Steven and {Morton}, Richard and {Del Zanna}, Giulio and {McIntosh}, Scott W. and {Karak}, Bidya Binay and {Gibson}, Sarah and {Samanta}, Tanmoy and {He}, Jiansen and {Chen}, Yajie and {Wang}, Linghua},
        title = "{Global maps of the magnetic field in the solar corona}",
      journal = {Science},
         year = 2020,
        month = aug,
       volume = {369},
       number = {6504},
        pages = {694-697},
          doi = {10.1126/science.abb4462},
archivePrefix = {arXiv},
       eprint = {2008.03136},
 primaryClass = {astro-ph.SR},
       adsurl = {https://ui.adsabs.harvard.edu/abs/2020Sci...369..694Y}
}

\end{document}